◆◆ Scientific
◆◆◆ Research

# Performance Study of Mobile TV over Mobile WiMAX Considering Different Modulation and Coding Techniques

**Jamil Hamodi[1,2], Ravindra Thool[1], Khaled Salah[3], Anwar Alsagaf[2], Yousef Holba[2]**

[1]Information Technology Department, SGGS Institute of Engineering and Technology, Nanded, India
[2]Computer Engineering Department, Hodeidah University, Hodeidah, Yemen
[3]Computer Engineering Department, Khalifa University of Science Technology and Research, Sharjah, UAE
Email: jamil_hamodi@yahoo.com, rcthool@yahoo.com, khaled.salah@kustar.ac.ae, Anwar_alsaqaf@hotmail.com,
yousef_holba2007@yahoo.com

## ABSTRACT

**With the advent of the wide-spread use of smart phones, video streaming over mobile wireless networks has suddenly taken a huge surge in recent years. Considering its enormous potential, mobile WiMAX is emerging as a viable technology for mobile TV which is expected to become of key importance in the future of mobile industry. In this paper, a simulation performance study of Mobile TV over mobile WiMAX is conducted with different types of adaptive modulation and coding taking into account key system and environment parameters which include the variation in the speed of the mobile, path-loss, scheduling service classes with the fixed type of modulations. Our simulation has been conducted using OPNET simulation. Simulation results show that dynamic adaptation of modulation and coding schemes based on channel conditions can offer considerably more enhanced QoS and at the same time reduce the overall bandwidth of the system.**

## KEYWORDS

## 1. Introduction

The primary challenge for present and future communication systems is the ability to transport multimedia content over a variety of networks in an energy efficient manner at different channel conditions and bandwidth capacities with various requirements of QoS [1]. Worldwide Interoperability for Microwave Access (WiMAX) technology is one of the future communications capable of offering high QoS at high data rates for IP networks. The high data rate and Quality of Service (QoS) assurance provided by this standard has made it commercially viable to support multimedia applications such as video telephony, video gaming, and mobile TV broadcasting. System architecture to support high definition video broadcasting (like MPEG-X, H.264/AVC and SVC) that provides a mobility of 30 kmph in an urban and sub-urban environment has been developed [2,3].

Mobile TV is a technology that enables users to transmit and receive TV program data through IP-based wired and wireless networks. Users can enjoy IPTV services anywhere and anytime with mobile devices. Four kinds of mobile TV technology approach, namely: mobile TV over IP that was discussed in this work, IPTV over mobile device, cellular IPTV, internet IPTV. However, with rapid adaptation to the user's requirements, Mobile TV may be preeminent in the future. It is convenient for users who can access an IPTV service through various wireless networks with mobile devices.

In cellular communication, the quality of a signal received by user equipment depends on a number of factors, namely: the distance between the desired user and interfering base stations, path loss exponent, log-normal shadowing, short term Rayleigh fading and noise. In order to improve system capacity, peak data rate and coverage reliability, the signal transmitted by a particular user is modified to account for the signal quality varia-tion through a process commonly referred to as link adaptation. Traditionally, CDMA systems have used fast power control as the preferred method for link adaptation. Recently, Adaptation Modulation and Coding (AMC) has offered an alternative link adaptation method that promises to raise the overall system capacity. Based on channel conditions, the scheme can be changed on per frame and per user basis. In order to maximize throughput in a channel varying in time, Adaptive and Modulation Coding can be used [4]. AMC provides the flexibility to match the modulation-coding scheme to the average channel conditions for each user. With AMC, the power of the transmitted signal hold constant over a frame interval, and the modulation and coding format are changed to match the current received signal quality or channel conditions. In a system with AMC, users close to the BS are typically assigned higher order modulation with higher code rates (e.g. 64 QAM with R = 3/4 turbo codes), but the modulation-order and/or code rate will decrease as the distance from BS increases [5].

Our work in this paper is geared towards investigating the performance study of Mobile TV (VoD) over Mobile WiMAX networks when considering different adaptive modulation and coding schemes using simulation software OPNET Modeler. OPNET provides comprehensive development of network models including all the necessary parameters that need to be reflected in the design procedure of PHY and/or MAC layers. A series of simulation scenarios under OPNET for broadband wireless communication are developed. In sharp contrast to prior work [6], the research work and results presented in this paper focus mainly on the use of real-time audio/video movies coded by MPEG-4 for modeling and simulation Mobile TV deployment over Mobile WiMAX. This paper, in the main, aims to establish a comparative study of performance for Mobile TV (VoD) over Mobile WiMAX under varying speeds of the mobile nodes and using different path-loss models under both AMC and fixed types of modulation techniques, and to identify the factors affecting the video performance.

The rest of this paper is organized as follows: Section 2 presents a brief overview of WiMAX. Section 3 presents performance metrics of video over WiMAX. Video streaming over WiMAX is given in Section 4. Section 5 describes the practical steps to be taken prior to simulation. Simulation results and analysis obtained are provided in Section 6. Section 7 is naturally the conclusion of my findings as a whole and a summation of my modest research endeavor.

## 2. Background and Preliminaries

This section provides relevant background and preliminaries pertaining to QoS support and the physical layer on WiMAX modulation and coding schemes. The Wi-

MAX physical layer is based on orthogonal frequency division multiplexing (OFDM). OFDM is the transmission scheme of choice to enable high-speed data, video, and multimedia communications and is used by a variety of commercial broadband systems, including DSL, Wi-Fi, Digital Video Broadcast-Handheld (DVB-H), and MediaFLOS, besides WiMAX. The standard defined within OFDM is an elegant and efficient scheme for high data rate transmission in a non-line-of-sight or multipath radio environment. Five PHY layer interfaces are in the IEEE 802.16 standard. These physicals interfaces are Wireless MAN-SC is known as SC operating in LOS condition at a frequency between 10 - 66 GHz, Wireless MAN-SC is known as SCa operating in NLOS condition at a frequency below 11 GHz, Wireless MAN-OFDM known as OFDM operating in NLOS condition at a frequency below 11 GHz, Wireless MAN-OFDMA known as OFDMA operating in NLOS condition at a frequency below 11 GHz, and last PHY interfaces is a Wireless HUMAN with frequency below 11 GHz. Among these PHY interface standards Wireless-MAN OFDM, a 2.5 GHz base frequency using 5 MHz has been used in our simulation that provides a 512-point TDD-based OFDM PHY for point-to-multipoint operations in NLOS conditions on frequencies between 2 GHz and 11 GHz [7].

WiMAX supports a variety of modulation and coding schemes and allows the scheme to change on a burst-by-burst basis per link, depending on channel conditions. Using the channel quality feedback indicator helps the mobile provide the base station with feedback on the downlink channel quality. For the uplink, the base station can estimate the channel quality based on the received signal quality. The base station scheduler can take into account the channel quality of each user's uplink and downlink and assign a modulation and coding scheme that maximizes the throughput for the available signal-to-noise ratio. Adaptive modulation and coding significantly increases the overall system capacity, as it allows real-time trade-off between throughput and robustness on each link. The modulation used in WiMAX is the orthogonal frequency division multiplexing (OFDM). WiMAX OFDM features multiple subcarriers ranging from a minimum of 256 up to 2048, each modulated with QPSK, 16 QAM, or 64 QAM modulation, where 64 QAM is optional on the uplink channel. The advantage of orthogonality is that it minimizes self-interference, a major source of error in receiving signals in wireless communications.

Channel coding schemes are used to help reduce the SINR requirements by recovering corrupted packets that may have been lost due to burst errors or frequency selecting fading. These schemes generate redundant bits to accompany information bits when transmitting over a channel. Coding schemes include convolution coding (CC) at various coding rates (1/2, 2/3 and 3/4) as well as conventional turbo codes (CTC) as various coding rates (1/2, 2/3, 3/4, and 5/6). The coding rate is the ratio of the encoded block size to the coded block size. The available coding rates for a given modulation scheme with the minimum signal to noise rate and the peak UL and DL data rates for 5 MHz channel mobile WiMAX with different information bits/symbol are listed in **Table 1** [8,9].

The key feature of adaptive modulation is that it increases the range over which a higher modulation scheme can be used, since the system can flex to the actual fading conditions, as opposed to having a fixed scheme that is budgeted for the worst case conditions [10]. Two sets of AMC schemes, AMC-1 and AMC-2, are considered to be used in DL in our simulation. These AMC profiles are same as in [11,12]. Each AMC is basically characterized by two threshold parameters, one is mandatory and the other is the minimum entry threshold for different modulation schemes. The mandatory exit threshold is the SINR at or below where this burst profile can no longer be used and where a change to a more robust (but also less frequency-use efficient) burst profile is required and the minimum entry threshold is the minimum SINR required to start using this burst profile when changing from a more robust burst profile [13]. **Table 2** shows the AMC profile used in our simulation. Here, AMC-2 is conservative AMC as it is using lower order MCS most of the time.

WiMAX supports different signal bandwidths ranging from 1.25 to 20 MHz to facilitate transmission over longer ranges in different multipath environments. In wireless communication systems, information is transmitted between the transmitter and the receiver antenna by electromagnetic waves. During propagation, electromagnetic waves interact with the environment, thereby causing a reduction of signal strength. Another limiting factor for higher sustained throughput in wireless communications, especially when the terminal nodes have the mobility, is caused by reflections between a transmitter and receiver, viz, a propagation path between the transmitter and the receiver is regarded. The propagation path between the transmitter and the receiver may vary from simple line-of-sight (LOS) to very complex one due to diffraction, reflecting and scattering [14]. To estimate the performance of mobile TV over WiMAX channels, propagation models in [6] are often used. Path loss models describe the signal attenuation between a transmitting and a receiving antenna as a function of the propagation distance and other parameters which provide details of the terrain profile required to estimate the attenuating signals. Path loss models represent a set of mathematical equations and algorithms which apply to radio signal propagation prediction in certain environments [15]. Path-loss is highly dependent on the propagation model, the common propagation models namely Free Space, Suburban Fixed (Erceg), Outdoor to Indoor and Pedestrian Environment and Vehicular Environment are given in **Table 3**. These models are used in mobile WiMAX performance evaluation through OPNET simulation.

**Table 1.** Mobile WiMAX PHY data rate and SINR for 5 MHz channel.

| Modulation scheme | Coding | Information bits/symbol | Minimum SINR (dB) | Dwon-link rate (Mbps) | Up-link rate (Mbps) |
|---|---|---|---|---|---|
| QPSK | 1/2 | 1 | 5 | 3.17 | 2.28 |
| | 3/4 | 1.5 | 8 | 4.75 | 3.43 |
| 16 QAM | 1/2 | 2 | 10.5 | 6.34 | 4.57 |
| | 3/4 | 3 | 14 | 9.5 | 6.85 |
| 64 QAM | 1/2 | 3 | 16 | 9.5 | 6.85 |
| | 2/3 | 4 | 18 | 12.6 | 9.14 |
| | 3/4 | 4 | 20 | 14.26 | 10.28 |

**Table 2.** AMC profiles.

| Modulation and coding schemes | AMC-1 | | AMC-2 | |
|---|---|---|---|---|
| | Mandatory exit threshold (dB) | Minimum entry threshold (dB) | Mandatory exit threshold (dB) | Minimum entry threshold (dB) |
| QPSK 1/2 | −20 | 2.0 | −20 | 2.0 |
| QPSK 3/4 | 5.0 | 5.9 | 11 | 11.9 |
| 16 QAM 1/2 | 8.0 | 8.9 | 14 | 14.9 |
| 16 QAM 3/4 | 11 | 11.9 | 17 | 17.9 |
| 64 QAM 1/2 | 14 | 14.9 | 20 | 20.9 |
| 64 QAM 2/3 | 17 | 17.9 | 23 | 23.9 |
| 64 QAM 3/4 | 19 | 19.9 | 25 | 25.9 |

IEEE 802.16 MAC defines up to five separate service classes to provide QoS for various types of applications. The service classes include: Unsolicited Grant Scheme (UGS), Extended Real Time Polling Service (ertPS), Real Time Polling Service (rtPS), Non Real Time Polling Service (nrtPS) and Best Effort Service (BE). Each service class has its own QoS parameters such as the way to request bandwidth, minimum throughput requirement and delay/jitter constraints. These service classes are [16] [17] described as follows: **UGS** is designed for constant bit rate (CBR) real-time traffic flows generating fixed-size data packets such as E1/T1 circuit emulation. It provides a fixed periodic bandwidth allocation. Once the connection is setup, there is no need to send any other requests. The main QoS parameters are maximum sustained rate, maximum latency and tolerated jitter (the maximum delay variation). **ertPS** is designed to support VoIP with silence suppression. No traffic is sent during the silent periods. ertPS service is similar to UGS in that the BS allocates the maximum sustained rate in active mode, but no bandwidth is allocated during the silent period. There is a need to have the BS poll the MS during the silent period to determine if the silent period has ended. The QoS parameters are the same as those in UGS. **rtPS** is for variable bit rate (VBR) real-time traffic such as MPEG compressed video. Unlike UGS, rtPS bandwidth requirements vary and so the BS needs to regularly poll each MS to determine what allocations needed to be made. The QoS parameters are similar to the UGS, but minimum reserved traffic rate and maximum sustained traffic rate need to be specified separately. For UGS and arts services, these two parameters are the same, if present. **nrtPS** is for non-real-time VBR traffic with no delay guarantees. Only the delay minimum rate is guaranteed. File Transfer Protocol (FTP) traffic is an example of applications using this service class. **BE** (most of data traffic falls into this category) guarantees neither delay nor throughput. The bandwidth will be granted to the MS if and only if there is a leftover bandwidth from other classes. In practice most implementations allow specifying minimum reserved traffic rate and maximum sustained traffic rate even for this class.

## 3. Video Performance Metrics

This section discusses some issues related to performance metrics of video transmission. Performance metrics can be classified as objective and subjective quality measures. Objective measures that observe packet transmissions include packet loss, packet delay, packet jitter, and traffic load throughput rates. Other objective metrics that attempt to quantify video quality perceptions include the ITU video quality metric (VQM) and peak signal to noise ratio (PSNR), which measure the codec's quality of reconstruction. Subjective video quality is a subjective characteristic of video quality. It is concerned with how video is perceived by a viewer and designates his or her opinion on a particular video sequence. The main idea of measuring subjective video quality is the same as in the

**Table 3. Path-loss models.**

| Propagation Model | Mathematical formulation | Description |
|---|---|---|
| Free space model | $P_{rx}(r) = P_{tx} G_{rx} G_{tx} / \left( (4\pi)^2 r^2 L \right)$ , $P_{rx}$ and $P_{tx}$ are received power in watts, respectively; $G_{rx}$ and $G_{tx}$ are the gain of the receiving and transmitting antennas, respectively; $L$ is the system-loss factor. | It is a mathematical model hardly applicable without considering the fading effect due to multi-path propagation. |
| Erceg's suburban fixed model | $PL = H + 10\gamma \log_{10}(d/d_0) + X_f + X_h + s$ , $PL$ is the instantaneous attenuation in dB, $H$ is the intercept and is given by free space path-loss at the desired frequency over a distance of $d_0 = 100$ m. $\bar{a}$ is a Gaussian random variable over the population of macro cells within each terrain category. $X_f$ and $X_h$ are the correlation factors of the model for the operating frequency and for the MS antenna height, respectively | It is based on extensive experimental data collected at 1.9 GHz in 95 macro cells of suburban areas across the United States. Very large cell size, base stations with high transmission power and higher antenna height. Subscriber stations are of very low mobility |
| Outdoor-to-indoor and pedestrian path-loss environment | $PL = 40 \log_{10} R + 30 \log_{10} f + 49$ . $PL$ is the instantaneous attenuation in dB, $R$ is the distance between the base station and the mobile station in kilometers and $f$ is the carrier frequency | Small cell size, base stations with low antenna heights and low transmission power are located outdoors while pedestrian users are located on streets and inside buildings and residences |
| Vehicular environment | $PL = 40\left(1 - 4*10^{-3} * \Delta h_b\right) \log_{10} R - 18 \log_{10} \Delta h_b - 21 \log_{10} f + 80 dB$ . $R$ is the distance between the base station and the mobile station, $f$ is the carrier frequency and $\Delta h_b$ is the base station antenna height in meters | Larger cells and higher transmitter power. All subscriber stations have a high mobility |

 

Mean Opinion Score (MOS). QoS requirement is very important for deploying IPTV and VoD as real time services over WiMAX networks. In order to assess the performance of video transmission systems, a suite of relevant performance metrics was identified to appropriately benchmark the system. Video on demand (VoD) deployments over WiMAX is affected by time varying bandwidth, packet delays, and losses. Since users expect high service quality regardless of the underlying network infrastructure, a number of metrics were collectively used to measure the video content streaming performance to ensure compliance and user's quality of experience (QoE) [18]. The following objective measures, widely used in the video content performance analysis employed, are packet loss ratio (PLR), packet delay (PD), packet jitter, and minimum throughput. The performance parameters affecting video have been shown in **Table 4** as in [19].

## 4. Video Streaming over Wimax

To provide a proper IPTV service to end users, this section describes the architecture of IPTV over WiMAX. IPTV providers must have an appropriate IP network to guarantee QoS at the services level. Our IPTV architecture over WiMAX network split into five subsystems, as can be seen in **Figure 1**: Video content is the first subsystem of our model, in this subsystem the servers store video content of any type of movies and audio content source. Different video types store the video source as national TV broadcasters, local broadcasters, Internet TV operations and any other future video broadcast service. Our video content held in this part has been variable bit rate videos encoded. The video content is delivered to the WiMAX network through the long distance, high capac-ity content distribution core network. The core network distributes the video flows from the header to the distribution network of the service provider. The distribution network goes from the end of the core network to the aggregation router, where the access network starts. The access network lets the user connect to the service provider and allows access to the multimedia content. The first requirement of an access network is to have enough bandwidth to support multiple IPTV channels for each subscriber, while it allows other services (telephony and data). Finally, the customer network (mobile receivers) enables communication and information exchange between the computers and devices connected to the services offered by a service provider.

There are some of researches work have explored WiMAX in the context of real-time and stored video applications. An in-depth performance evaluation of mobile WiMAX is carried out using adaptive modulation and coding under the real-like simulation environment of OPNET presented in [6]. They have evaluated the per-formance parameters of mobile WiMAX with respect to different modulation and coding schemes. Their performance has been evaluated in terms of average throughput, average data-dropped, the MOS value of voice application and the BW usage in terms of UL data burst usage when deployed VoIP on WiMAX Networks. They have been observed that using lower order modulation and coding schemes, the system provides better performance in terms of throughput, data dropped and MOS at the cost of higher BW usage.

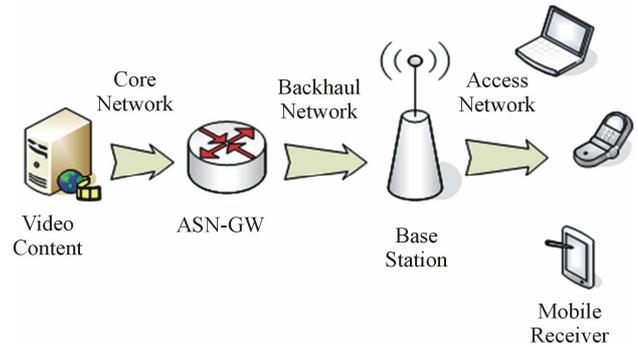

**Figure 1.** Mobile TV over WiMAX architecture.

**Table 4.** Performance parameters for deploying VoD.

| Metrics | Mathematical formulation | Description | Acceptable |
|---|---|---|---|
| Packet loss ratio (PLR) | $PLR = \left( \dfrac{\text{lost}_{packet}}{\text{lost}_{packet} - \text{received}_{packet}} \right)$ | **PLR** is the corrupted, lost, or excessively delayed packets divided by the total number of packets expected at the video client station. | $10^{-3}$ |
| Packet End-to-End Delay (E2E) (ms) | $D_{E2E} = Q\ (d_{proc} + d_{queue} + d_{trans} + d_{prop})$, where: $Q$ is the number of network elements between the media server and mobile station. $d_{proc}$ is the processing delay at a given network element. $d_{queue}$ is the queuing delay at a given network element. $d_{trans}$ is the transmission time of a packet on a given communication link between two network elements. $d_{prop}$ is the propagation delay across a given network link | **Packet delay** is the average packet transit time between the media server and the video client station. | <400 |
| Packet delay variation (PDV) or packet jitter (ms) | $j_{pkt} = t_{actual} - t_{expected}$, where: $t_{actual}$ is the actual packet reception time. $t_{expected}$ is the expected packet reception time. | **Packet jitter** is defined as the variability in packet delay within a given media at video client station. | <50 |
| Throughput (bps) | The **throughput** for variable bit rate (VBR) traffic loading is dynamic in nature and it is a function of the scene complexity and associated audio content. Variable bit rate (VBR) traffic loads is typically quoted as peak throughput ranges. | **Throughput** is defined as the traffic load that the media stream will add to the network. It can be measured in bits/sec. | [221-5311] |

Challenges for delivering IPTV over WiMAX are discussed in [20]. These include the challenges for QoS requirements. Also, they describe the transmission of IPTV services over WiMAX technology, and the impact of different parameters in WiMAX network when deploying this service. An intelligent controller has been designed based on fuzzy logic to analyze QoS requirements for delivering IPTV over WiMAX. Also, an intelligent controller based on fuzzy logic in [21] is used to analyze three parameters: jitter, losses and delays that affect the QoS for delivering IPTV services. The aim is to define a maximum value of link utilization among links of the network.

An OPNET Modeler is used to design, characterize, and compare the performance of video streaming to WiMAX and ADSL. The simulation results indicate that, ADSL exhibits behavior approached the ideal values for the performance metrics while WiMAX demonstrates promising behavior within the bounds of the defined metrics. The work in [18] is extended work in [22] to include generation and integration of a streaming audio component, also enhances the protocol stack to include the real time protocol (RTP) layer. Network topology is redesigned to incorporate WiMAX mobility. Also, they include characterization of WiMAX media access control (MAC) and physical (PHY) layer. Simulation scenarios are used to observe the impact on the four performance metrics. Also, in [23] simulation is used to compare the performance metrics between ADSL and WiMAX by varying the attributes of network objects such as traffic load and by customizing the physical characteristics to vary BLER, packet loss, delay, jitter, and throughput. Simulation results demonstrate considerable packet loss. ADSL exhibits considerably better performance than the WiMAX client stations.

However, many of recent works explore the performance studies of WiMAX under different modulation and coding schemes. For example, Telagarapu *et al.* [24] analyzed the physical layer of WiMAX with different modulation techniques like BPSK, QPSK, QAM and comparison of QPSK modulation with and without Forward Error Correction methods. Singh *et al.* [25] provided an analysis of the Bit error rate Vs SNR and Block error rate Vs SNR curves using QPSK, 16 QAM and 64

QAM in an AWGN channel with parameters analyzed for different code rates. It was found that the greater number of symbols transmitted per block, the transmission quality decreases.

Islam *et al.* [26] evaluated WiMAX system under different combinations of digital modulation (BPSK, QPSK, 64-QAM and 16-QAM) and different communication channels AWGN and fading channels (Rayleigh and Rician), and the WiMAX system incorporates Reed-Solomon (RS) encoder with Convolutional encoder with 1/2 and 2/3 rated codes in FEC channel coding. Also the work in [27] evaluated and analyzed the bit error rate performance of WiMAX Physical layer with the implementation of different concatenated channel coding schemes under various digital modulations over realistic channel conditions. Simulation results demonstrate the out performance of the concatenated Cyclic Redundancy Check and Convolutional (CRC-CC) code when compared to Reed-Solomon and Convolutional (RS-CC) code and the proposed WiMAX system achieves good error rate performance under QAM modulation technique in AWGN, Rayleigh and Rician fading channels. Whereas Rehman *et al.* [28] used the Simulink tool to evaluate WiMAX system under different combinations of digital modulation and different communication channels AWGN and fading channels. Both Reed Solomon (RS) encoder with Convolutional encoder with 1/2 and 2/3 rated codes in FEC channel coding are incorporated in WiMAX system.

## 5. Simulation Model

The network topology for our test-bed network is given in **Figure 2**. The simulation model of our case study network deployed with 7-Hexagonal celled WiMAX, with multiple subscriber stations in the range of a base station. The base stations are connected to the core network by an IP backbone. The IP backbone is connected to the server backbone via an ASN-GW gateway to support the mobility in the WiMAX network. The ASN-GW gateway, IP backbone, and server backbone together represent the service provider company network. We have uses only one SS with varying speed. This node is (mobile 5_1) move along the trajectories indicated by white color around cells. The green bidirectional dotted lines represent the generic routing encapsulation (GRE) tunnels. The common attributes used for network configuration are highlighted in **Table 5**.

Video streaming over wireless networks is a challenging task. This is due to the high bandwidth required and the delay sensitive nature of video more than most other types of application. As a result, a video trace of 2-hour MPEG-4 Matrix III movie is used in our simulation. This video coded by MPEG-4 part 2, is obtained from Arizona State University [29] with 352 × 288 frame resolution,

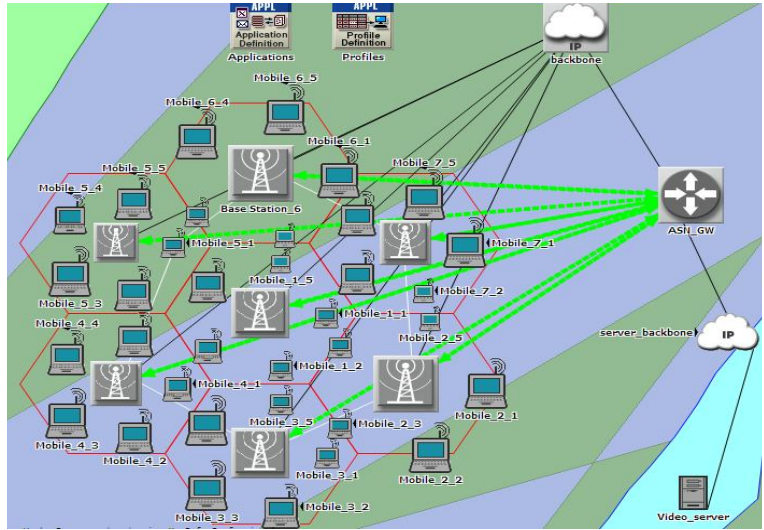

**Figure 2.** Network model of mobile TV over WiMAX.

<table>
<tr><td colspan="2" align="center">**Table 5.** Network configuration details.</td></tr>
<tr><td align="center">**Network**</td><td align="center">Mobile WiMAX Network</td></tr>
<tr><td align="center">**Cell Radius**</td><td align="center">0.2 Km</td></tr>
<tr><td align="center">**No. of Base Stations**</td><td align="center">7</td></tr>
<tr><td align="center">**No. of Subscriber Stations**</td><td align="center">5</td></tr>
<tr><td align="center">**IP Backbone Model**</td><td align="center">IP32_cloud</td></tr>
<tr><td align="center">**Video Server Model**</td><td align="center">PPP_sever</td></tr>
<tr><td align="center">**Link Model (BS-Backbone)**</td><td align="center">PPP_DS3</td></tr>
<tr><td align="center">**Link Model (Backbone-server Backbone)**</td><td align="center">PPP_SONET_OC12</td></tr>
<tr><td align="center">**Physical Layer Model**</td><td align="center">OFDM 5 MHz</td></tr>
<tr><td align="center">**Traffic Type of Services**</td><td align="center">Streaming Video</td></tr>
<tr><td align="center">**Application**</td><td align="center">Real Video streaming</td></tr>
<tr><td align="center">**Scheduling**</td><td align="center">rtPS</td></tr>
</table>

**Table 6.** Video codec traces characteristics.

| Parameters | Matrix III |
|---|---|
| Codec | MPEG-4 Part 2 |
| Frame Compression Ratio | 47.682 |
| Min Frame Size (Bytes) | 8 |
| Max Frame Size (Bytes) | 36,450 |
| Mean Frame Size (Bytes) | 3189.068 |
| Peak Frame Rate (Mbps) | 7.290 |
| Mean Frame Rates (Mbps) | 0.637 |
| Frame Rate (frames/seconds) | 25 |

and an encoding rate of 25 frames per seconds (fps). **Table 6** shows the mean and peak rates for this video code, and reflects only video frames. Our proposed simulation model traces driven using video and audio traces from the Matrix III movie while these traces reflect the individual frame sizes rather than the actual frame data. So, this work also adds audio frames, which is 21.6 as in [18].

Two independent instances of the video conferencing application are used to stream the separate and distinct video and audio components of the Matrix III movie. These two applications configured to work simultaneously stream in the profile configuration. The key parameters of this application configuration are the frame inter-arrival time and frame size. The incoming inter-arrival times are configured to the video and audio frame rates of 25 and 21.6, respectively. It should be noted that the outgoing inter-arrival time remains set to "none" in order to achieve unidirectional streaming from the video server. Furthermore, the frame size parameters are configured to explicitly script the video and audio traces.

## 6. Results and Discussion

The ninety nine scenarios were simulated and the results are collected and summarized in three cases depending on varying speed of the mobile node, varying path loss models, and for different types of service classes. For each case, the types of modulation and coding schemes are choosing one at a time to obtain one set of simulation results for the different performance measures of packet loss, packet delay, packet jitter, and traffic load throughput.

### 6.1. Case 1: Mobile Node with Different Speed

This subsection shows the simulation results of twenty seven scenarios which had been simulated under this case. This case is investigated how mobile TV speeds perform using different modulation and coding schemes offered by Mobile WiMAX. This evaluated the performance parameters, namely: packet jitter, packet E2E delay, data drop, and throughput of the mobile node, and showed which the appropriate speed is the best for Mobile TV. The speed of the mobile is provided in kmph while path-loss model and scheduling service are kept constant. Path-loss model is chosen as free space and service class as rtPS is considered.

The average packet jitter, and average E2E delay with fixed and adaptive modulation have been shown in **Figures 3(a)** and **(b)**. **Figure 3(a)** shows the average variation of jitter for audio/video Mobile TV over Mobile WiMAX networks for different speed, video quality is best if the jitter is zero. Axis scale of **Figure 3(a)** is as log scale because the variation results are very small around zero. As shown in **Figure 3(a)**, average audio/video jitter is approximately zero for higher MCS and

AMC under different speeds which is about 2.5753E−05 seconds, whereas other MCS as QPSK has a more average variation of jitter. From the results in **Figure 3(a)**, it is observed that WiMAX using higher MCS (64 QAM 2/3, 64 QAM 3/4) and AMC (AMC-1, AMC-2) as a modulation technique shows better jitter compared with other MCS (QPSK, 16 QAM). Average End-to-End delay for different speeds under MCS shows in **Figure 3(b)**, as we can see the average E2E delay of different speeds give lower packet E2E delay for audio/video Mobile TV under higher modulation and coding MCS (16 QAM, 64 QAM) and AMC compared with QPSK.

Intuitively, increasing data drop is affected by the speed of SS. However, when the speed of SS increases, the hand-off frequency increases, which results in increased data dropped and thereby decreased throughput. As can be viewed from **Figure 4(a)**, the average data drop is significantly higher when SS moves with a greater speed as 150 kmph. The effect of data drop naturally decreases the average WiMAX throughput as shown in

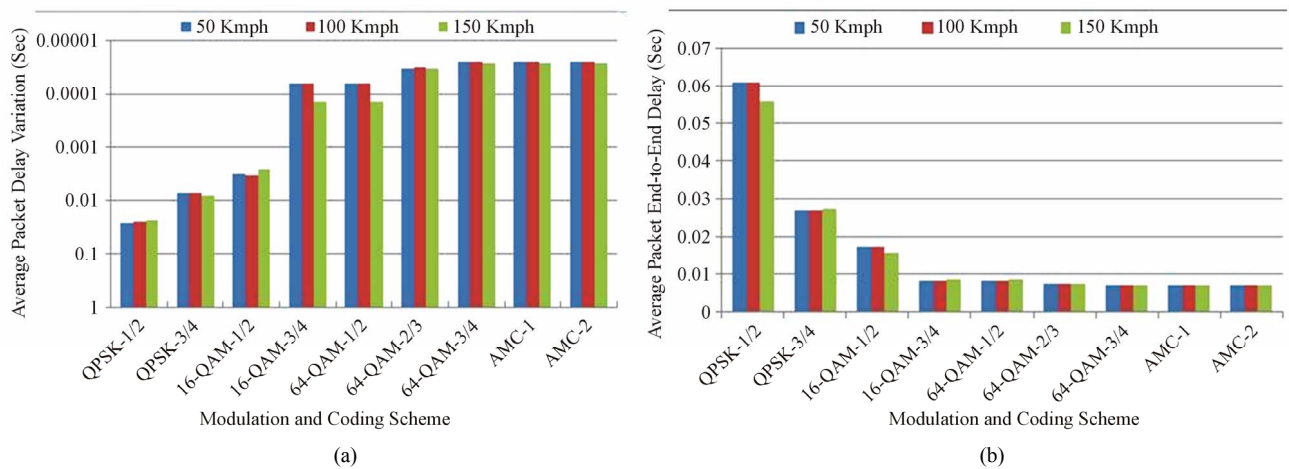

**Figure 3.** (a) Average video jitter; (b) Average packet End-to-End delay.

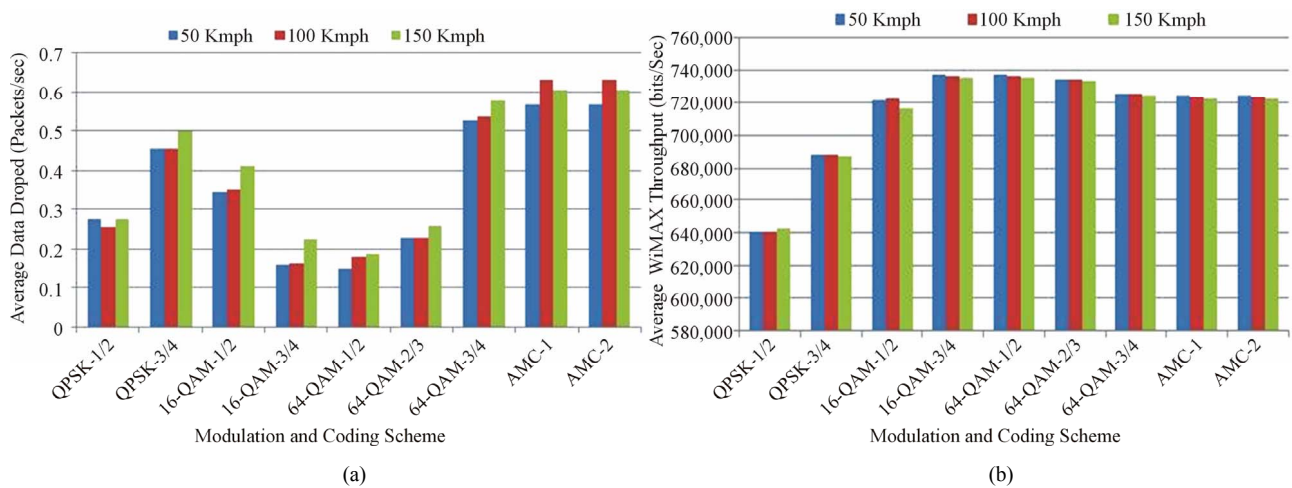

**Figure 4.** (a) Average packet data dropped for SS node; (b) Average WiMAX throughput for SS node.

Figure **4(b)**. From **Figure 4(a)** it is observed that the data dropped is very low for 16 QAM 3/4, and 64 QAM 1/2 modulation schemes and varies with speed. We know that a higher order modulation scheme is more sensible to SINR. As the SS is moving through the cell, it faces a different SINR value depending upon the distance from the BS and the propagation environment. With increasing distance, the SINR decreases and the higher order MCS gives more BLER than the lower order MCS for the same SINR value. For that the higher order 64 QAM 3/4 has more data dropped and AMC modulation schemes with varying speeds. **Figure 5** shows the BLER for different MCS under 150 Kmph, yellow line in **Figure 5** shows the 64 QAM 3/4 which has more BLER than other MCS. The average throughput is taken as a measure that will give the average of observed throughput right through the simulation. Thus, the higher order of MCS 16 QAM 3/4, 64 QAM 1/2, and 64 QAM 2/3 has the best performance (throughput) as observed in **Figure 4(b)**. According to the results as in **Figures 4(a)** and **(b)**, it is observed that WiMAX uses 64 QAM 1/2 and 16 QAM 3/4 as modulation techniques show better performance (high throughput) compared with other modulation techniques.

## 6.2. Case 2: Mobile Node with Different Path Loss

This subsection shows the simulation results of 36 scena-rios, performance parameters of each scenario observed for various modulation and coding schemes with respect to various path-loss models. We have considered in this case keeping the speed of SS constant as 50 Kmph and scheduling service classes as rtPS.

In this study fixed radius WiMAX networks are considered for all the path loss. As in our knowledge, outdoor to indoor and pedestrian path-loss model is designed for small and micro cell WiMAX network. For outdoor to indoor and pedestrian propagation model, as SS node that moves away from BS will encounter a significant drop in SINR and as the higher order MCS (such as 64 QAM 3/4) requires high value of SINR to give a good throughput; higher order MCS will face a very large amount of data drop as revealed from **Figure 6(a)**. However, outdoor to indoor and pedestrian has lower packet jitter and lower packet E2E delay as shown in **Figures 7(a)** and **(b)**. Path-loss of free space is lowest; hence, the reduction of SINR with the distance from BS is less which leads to better throughput, lower packet jitter, lower packet E2E delay, and lower packet data dropped from all varies MCS as shown in **Figures 6** and **7**. Again, as the reduction of SINR with distance from BS is less, the SS has to change its modulation scheme less frequently which results in very high throughput for AMC as shown in **Figure 6(b)**. Here, the suburban fixed model has been considered as hilly terrain with moderate to heavy tree density that representing rural environments and has highest path-loss. Also, same terrain is considered here in a vehicle model. As outdoor to indoor and pedestrian propagation model experiences a very high packet drop compared with the others, it gives the lowest throughput compared with other propagation model as can be observed from **Figure 6(b)**. For free space propagation model, 16 QAM, and 64 QAM has the same throughput under varied coding as in **Figure 6(b)**.

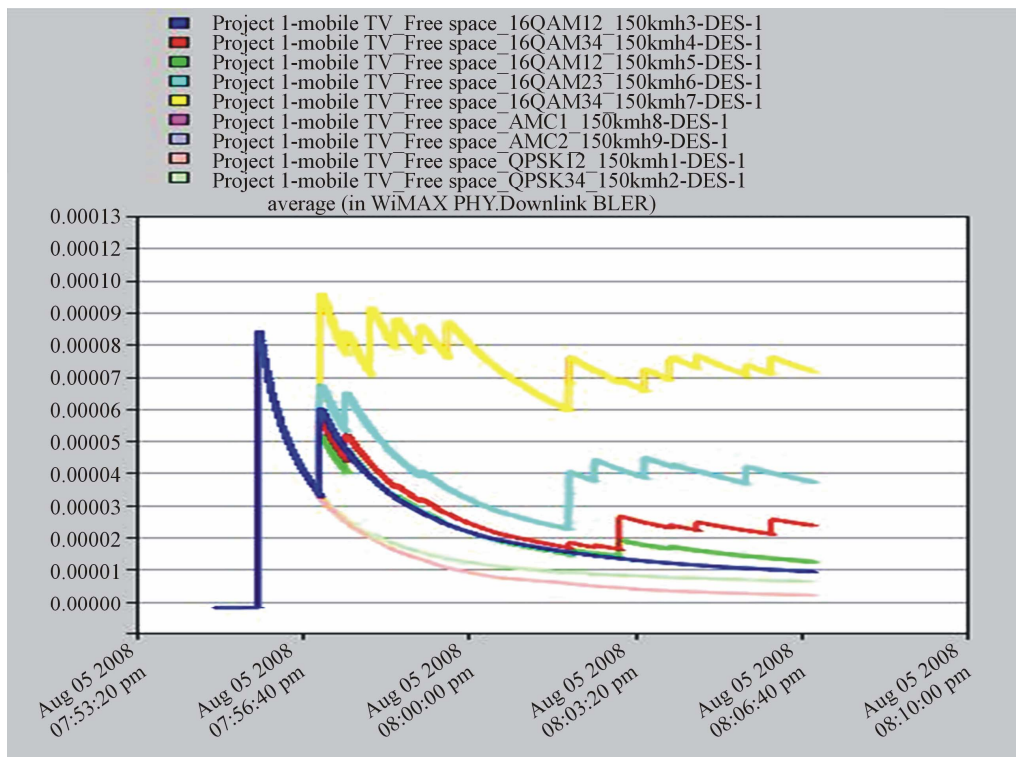

**Figure 5.** Average download BLER.

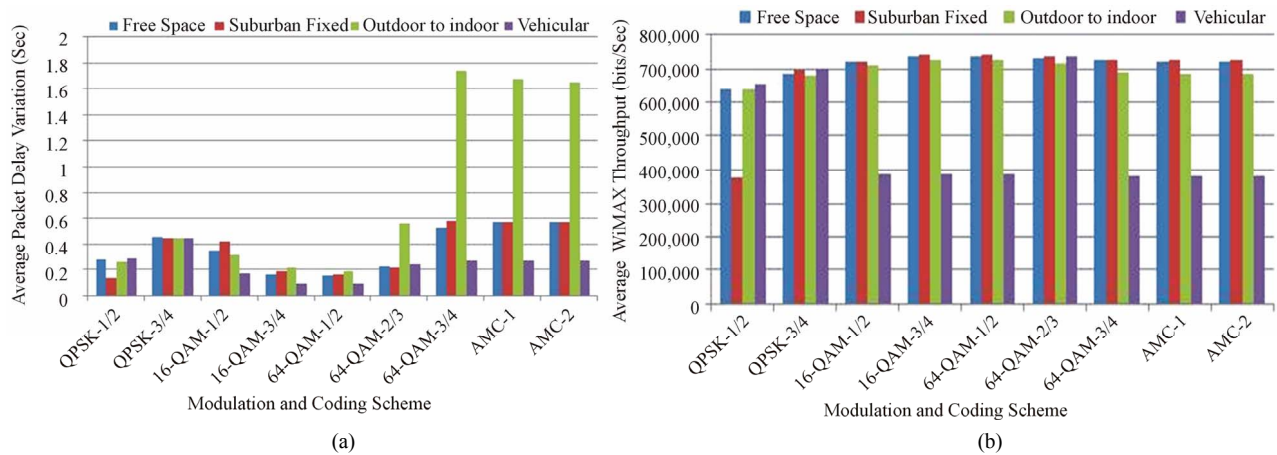

**Figure 6.** (a) Average packet data dropped for SS node; (b) Average WiMAX throughput for SS node.

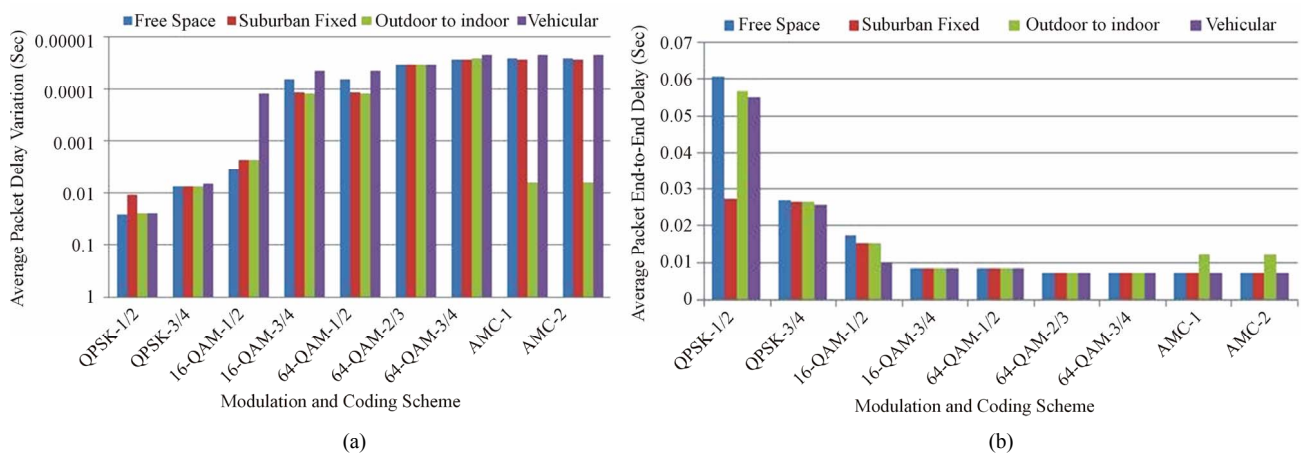

**Figure 7.** (a) Average video jitter; (b) Average packet End-to-End delay.

### 6.3. Case 3: Mobile Node with Different Classes

This subsection exhibits the simulation results of 36 scenarios which has been simulated under this case; speed and path loss kept constant. The speed of SS is chosen as 50 kmph while path-loss model as free space is considered. Different service classes are used in this case under various modulations and coding scheme to obtain the performance metrics: packet delay variation, packet End-to-End delay, data drooped for a mobile node, and WiMAX throughput for the mobile node.

To the best of the authors, UGS and ertps were designed to support VoIP. However, UGS is used for CBR,

whereas, A/V Mobile TV has VBR, it cannot use UGS in our simulation. So, UGS classes are not supported in our case study. On the other hand, ertps are designed to support real-time applications generating variable bit rate traffic, and are designed for VoIP service with activity detection. Results obtained shows that this class has worse performance in all metrics as shown in **Figures 8** and **9** it has lower throughput, higher jitter, and higher packet E2E delay. **Figures 8(a)** and **(b)** show packet delay jitter and packet E2E delay for different service classes: rtPS, nrtPS, ertPS, and BE. rtps, nrtps, and BE has given the best performance, which reveals that for higher modulation and coding, all service classes give equal throughput, equal packet jitter, and equal packet E2E delay. In **Figure 9(b)**, it can be observed that rtps classes' providers have greater throughput than other classes, nrtPS, and BE. The reason is that rtPS is designed for streaming Audio or Video.

## 7. Conclusion

This paper presents a comparative performance study of

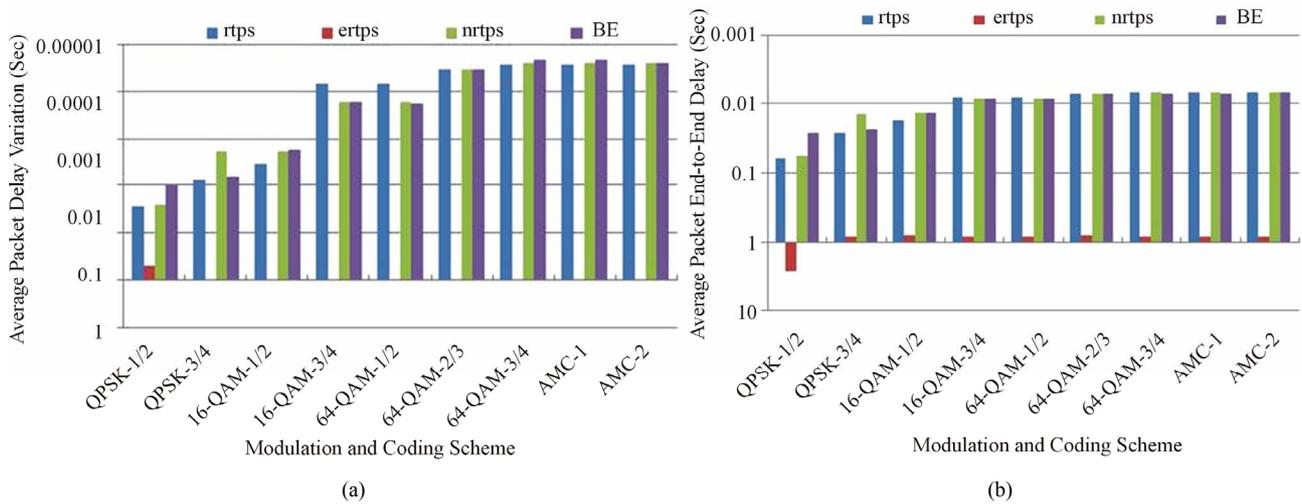

(a)

(b)

**Figure 8.** (a) Average video jitter; (b) Average packet End-to-End delay.

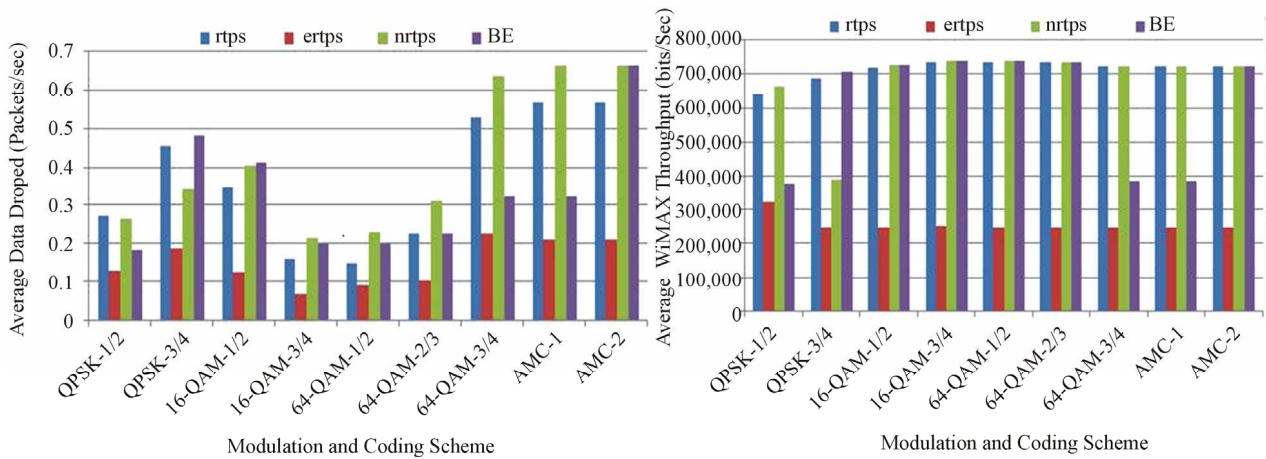

**Figure 9.** (a) Average packet data dropped for SS node; (b) Average WiMAX throughput for SS node.

Mobile TV over mobile WiMAX network with respect to different modulation and coding schemes under different parameters including speed of mobile node, path-loss models, and MAC service classes. The performance has been evaluated in terms of average packet jitter, average packet E2E delay, average throughput, and average data-dropped. It has been demonstrated, using OPNET simulation, that higher order modulation and coding schemes (namely, 16 QAM and 64 QAM) provide better performance. Also, it has been shown that AMC schemes (namely, AMC-1 and AMC-2) give relatively comparable performances to 64 QAM 3/4 scheme. In this work, simulation results show that the free space path loss is best for deploying A/V video application over different mobile node speeds, whereas "outdoor to indoor" is the worst case with the highest packet drop rate. Moreover, our simulation showed that rtPS scheduling service class is the most appropriate scheduling service for A/V video application.

## Acknowledgements

The authors thank Professor Hussein Omar Qadi, President of Hodeidah University for his supporting to develop the academic research and researchers in the Hodeidah University.